\documentclass[superscriptaddress,showkeys,twocolumn,nofootinbib,longbibliography]{revtex4-1}

\usepackage{amsmath}
\usepackage{amssymb}
\usepackage{graphicx}
\usepackage{epsf}
\usepackage{slashed}
\usepackage{enumitem}
\usepackage[czech,english]{babel}
\usepackage[cp1250]{inputenc}
\usepackage{hyperref}
\usepackage{xcolor}
\usepackage{amscd}

\usepackage{cancel}
\usepackage{comment}

\usepackage[only,llbracket,rrbracket,llparenthesis,rrparenthesis]{stmaryrd} % for comparison with 4 similar parenthesis symbols
\usepackage{accsupp} % for ensuring the right Unicode codepoint upon pasting

\newcommand{\refer}[1]{(\ref{#1})}
\newcommand{\diff}{\mathrm{d}}

\newcommand{\Exp}[1]{\mathrm{e}^{#1}}

\newcommand{\myphi}{\phi_{TT}}

\usepackage{bbm} %\mathbbm{1}, \mathbbmss{1}, \mathbbmtt{1}

\usepackage[nodayofweek]{datetime}
\newdateformat{mydate}{\twodigit{\THEDAY}{ }\shortmonthname[\THEMONTH], \THEYEAR}
\usepackage[normalem]{ulem}
\begin{document}

\title{Scattering of kinks in coreless potentials}

\author{Ond\v{r}ej Nicolas Karp\'{i}\v{s}ek}
\email{karponius(at)gmail.com}
\affiliation{
Institute of Physics, Silesian University in Opava, Bezru\v{c}ovo n\'am. 1150/13, 746~01 Opava, Czech Republic.
}

\author{Luk\'a\v{s} Rafaj}
\email{lukasrafaj(at)gmail.com}
\affiliation{
Institute of Physics, Silesian University in Opava, Bezru\v{c}ovo n\'am. 1150/13, 746~01 Opava, Czech Republic.
}

\author{Filip Blaschke}
\email{filip.blaschke(at)physics.slu.cz}
\affiliation{
Research Centre for Theoretical Physics and Astrophysics, Institute of Physics, Silesian University in Opava, Bezru\v{c}ovo n\'am. 1150/13, 746~01 Opava, Czech Republic.
}

\begin{abstract}
We explore the relevance of the central hill for a symmetric double-well potential and its impact on the scattering of kinks in a scalar field theory in (1+1)-dimensions. This region controls the inner core structure of the kink. We study how the disappearance of analyticity in this region of the potential affects the resonant features in $K\bar{K}$ scattering.
%\remark{Version V1, compiled: \mydate\today, \currenttime.}
\end{abstract}

%\keywords{Kinks, collective coordinate model, mechanization}

\maketitle

%\tableofcontents

%\raggedbottom

%%%%%%%%%%%%%%%%%%%%%%%%%%%%%%%%%%%%%%%%%%%%%%%%%%%%%%%%%%
%%%%%%%%%%%%%%%%%%%%%%%%%%%%%%%%%%%%%%%%%%%%%%%%%%%%%%%%%%
%%%%%%%%%%%%%%%%%%%%%%%%%%%%%%%%%%%%%%%%%%%%%%%%%%%%%%%%%%
%%%%%%%%%%%%%%%%%%%%%%%%%%%%%%%%%%%%%%%%%%%%%%%%%%%%%%%%%%
\section{Introduction}
\label{sec:I}
%%%%%%%%%%%%%%%%%%%%%%%%%%%%%%%%%%%%%%%%%%%%%%%%%%%%%%%%%%
%%%%%%%%%%%%%%%%%%%%%%%%%%%%%%%%%%%%%%%%%%%%%%%%%%%%%%%%%%
%%%%%%%%%%%%%%%%%%%%%%%%%%%%%%%%%%%%%%%%%%%%%%%%%%%%%%%%%%
%%%%%%%%%%%%%%%%%%%%%%%%%%%%%%%%%%%%%%%%%%%%%%%%%%%%%%%%%%

Classical field theory with non-linear field interactions often leads to the presence of solitons. These are particle-like, extended objects that are remarkably stable under the effects of perturbation or soliton-soliton interactions and lead to a richness of dynamical phenomena. When the underlying model has in addition non-trivial topology, there exist topological solitons, which are absolutely stable  \cite{Manton:2004tk, Rajaraman:1982is, Vilenkin:2000jqa, Shnir:2018yzp}. 

Among the conceptually and mathematically simplest topological solitons are the so-called \emph{kinks}. These particle-like objects in one spatial dimension (manifesting as strings and domain walls in higher dimensions) are described by a single, real, self-interacting scalar field, say $\phi$. A local, Lorentz invariant description is afforded via the Lagrangian density
\begin{equation}\label{eq:generic}
\mathcal{L} = \frac{1}{2}\partial_\mu \phi \partial^\mu \phi -V(\phi)\,,
\end{equation}
where the potential $V(\phi)$ encodes the self-interaction. The kinks are present as static solutions for any potential $V(\phi)$ that has multiple vacua, i.e. field values $\phi = v_a$ for which $V(v_a) = V^\prime(v_a) = 0$. 

The mathematical framework embodied in Eq.~\refer{eq:generic} is so simple that one would not expect the dynamics of kinks to be particularly complicated. Since kinks have been the object of scientific interest for many decades now, one could be tempted to guess that the overall dynamic picture of how kinks interact with themselves or with the environment is fully mapped out. Therefore, it is more surprising to learn that the actual state of affairs is still far from ideal.

Indeed, while the
interactions of kinks with other kinks and/or anti-kinks have
been numerically investigated since 70ties  \cite{Sugiyama:1979mi, Campbell:1983xu, Moshir:1981ja, Anninos:1991un, Belova:1985fg, Kudryavtsev:1975dj}, the quantitative and qualitative dynamical picture of the associated phenomena, such as bouncing, bion formation, the role of radiation, spectral wall phenomenon, dynamical generation of delocalized modes, etc., has been achieved in both classical and quantum settings (with various degrees of completeness) only recently \cite{Manton:2020onl, Manton:2021ipk, Adam:2021gat, Dorey:2011yw, Adam:2022mmm, Weigel:2013kwa, Dorey:2023izf, Adam:2023kel, Adam:2023qgx, Dorey:2023izf, Adam:2023kel, Adam:2022bus, Adam:2019xuc, Adam:2023qgx, Dorey:2017dsn, Evslin:2022xmp, Evslin:2023oub, Evslin:2023egm, Evslin:2024czr} (one can also read about the somewhat intricate history of investigations of kink-anti-kink collisions in \cite{2019arXiv190903128K} and \cite{Belova:1997bq}).

The source for all this complexity must be somehow intricately encoded into the potential $V(\phi)$. 

Indeed, the choice of the potential dictates the underlying dynamics with precarious exactness. For instance, the so-called sine-Gordon (sG) model $V(\phi) = 2\sin^2(\phi/2)$ is completely integrable, and the kink-anti-kink ($K\bar K$) and kink-kink $(KK)$ solutions (and many others) are known in a closed form. The collisions of kinks are completely elastic, exemplifying perhaps the simplest behavior across the spectrum of all models. 

On the other hand, $K\bar K$ collisions in the $\phi^4$ (or double-well) model with the potential $V(\phi) = \bigl(1-\phi^2\bigr)^2/2$, which is regarded as a canonical representative, are very rich in 
dynamical aspects, such as the fractal structure of bouncing windows, bion chimneys, etc. Moreover, these features are universal, in that they are present for generic choices of potentials unless special circumstances prevent them from occurring, such as integrability or absence of resonant modes that facilitate energy transfer mechanism. 

This is exemplified in the so-called $\phi^6$ model with $V(\phi) = \phi^2\bigl (1-\phi^2\bigr )^2/2$ in which the $K\bar K$ collisions are devoid of this fractal structure, while the collisions of anti-kinks with kinks ($\bar K K$) produce it \cite{Adam:2022mmm, Dorey:2011yw}. Multiple other potentials have been explored, such as $\phi^8$ potential $V(\phi) = \phi^4\bigl(1-\phi^2\bigr)^2/2$ in which kinks have long, polynomial tails \cite{Christov:2018wsa, Manton:2018deu, Gani:2015cda, Gani:2021ttg, Christov:2018ecz, Belendryasova:2017wad} and the parametrically-dependent Christ-Lee potential, i.e. $V(\phi) = \bigl(\varepsilon^2 +\phi^2\bigr)\bigl(1-\phi^2\bigr)^2/(2+2\varepsilon^2)$ that smoothly interpolates between $\phi^4$ and $\phi^6$ potentials \cite{Dorey:2023izf}, to name just a few.

\subsection{The shape of a kink}

The shape of a static kink is in one-to-one correspondence with the shape of the potential between the minima that the kink interpolates. This is most easily seen from the static equation of motion
\begin{equation}
\phi^{\prime\prime}(x) = V^\prime(\phi)\,,
\end{equation}
which is equivalent to an equation of motion for a particle under the influence of the upside-down potential $-V(\phi)$. The ``time'' is the $x$-coordinate and the particle starts at one of the maxima, corresponding to the ``left'' vacuum, $v_{\rm L}$, at $x = -\infty$ and rolls down towards the other maximum, $v_{\rm R}$, that is reached at $x = +\infty$ \cite{Rajaraman:1982is}.

\begin{figure}[htb!]
\begin{center}
\includegraphics[width=0.7\columnwidth]{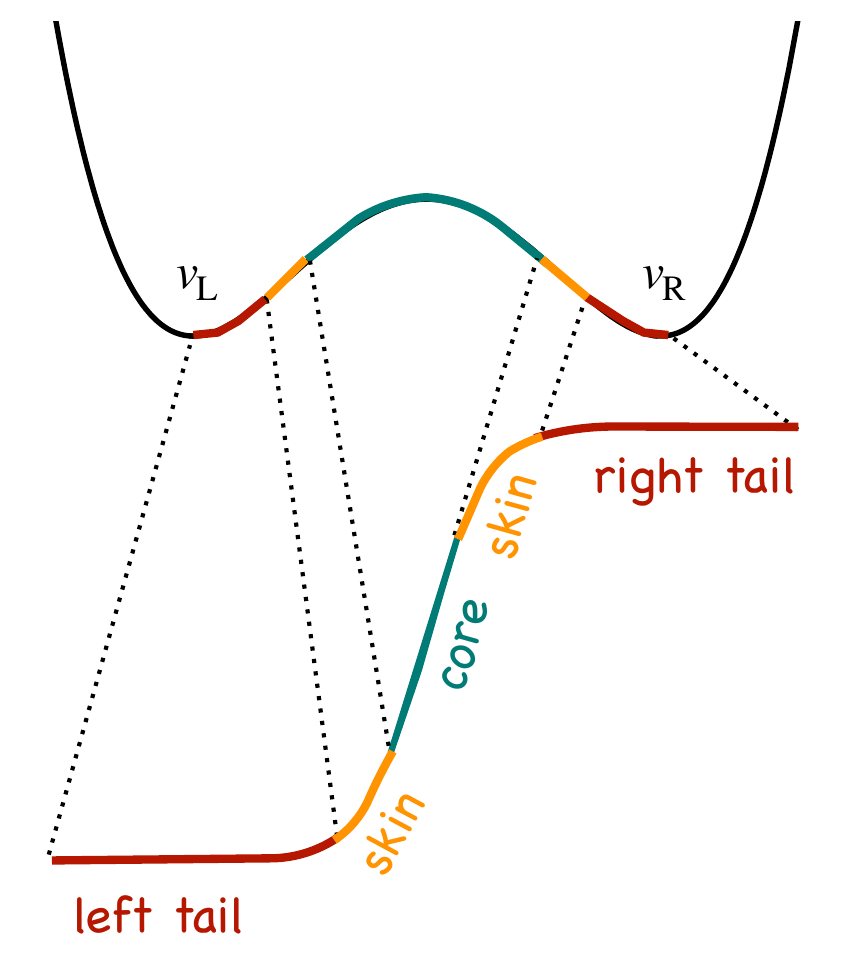}
\end{center}
{
    \caption{\small A pictograph of how the shape of a kink, interpolating between $v_{\rm L}$ and $v_{\rm R}$ vacua is affected by different parts of the potential.}
    \label{fig:potdata}
}
\end{figure}

For a generic potential with two vacua, and no other local minima, such as the one depicted in Fig.~\ref{fig:potdata}, we can talk about three distinct regions with qualitatively different impacts on the shape of the kink. 

The regions near vacua encode the \emph{tails} of the kink. Indeed, for non-zero curvature, i.e. $V^{\prime\prime}(v_{\rm L,R}) \equiv m_{\rm L,R}^2 >0$, the kink approaches vacua exponentially fast as $|\phi-v_{\rm L,R}| \sim \exp\bigl(\pm m_{\rm L,R}x\bigr)$. On the other hand, if $m = 0$, the tails are polynomial \cite{Christov:2018ecz, Christov:2018wsa}.

The region near the maximum controls the shape of the very center of the kink, which we shall call its \emph{core}. Denoting $\alpha^2 \equiv - V^{\prime\prime}(\phi_{\rm max})$, the kink's profile at its core (where also the most energy is concentrated and is identified with the kink's position) will be approximately given by a combination of $\sin(\alpha x)$ and $\cos(\alpha x)$. On the other hand, if $\alpha=0$, such is the case for $\phi^8$ potential mentioned above, the core's profile can be approximated by Jacobi functions, i.e. solutions to $\delta\phi^{\prime\prime} = 2 \delta \phi^3$.

Lastly, the regions near inflection points dictate what we subsume into the notion of the kink's \emph{skin}. Around the skin, the kink behaves as a quadratic function of $x$.\footnote{This is also true around a generic point where the first derivative is non-vanishing.}

We can be more precise and place the above notions onto a firm ground by defining the tails, core, and skin of a kink in a perturbative sense. 

Expanding the field around the vacuum, i.e. $\phi = v + \delta\phi$ and putting it into the equation of motion, we get 
\begin{equation}
\delta\phi^{\prime\prime} = m^2 \delta \phi + \frac{1}{2}V^{\prime\prime\prime}(v) \delta\phi^2 + \ldots
\end{equation}
At the leading order, the above equation is solved as $\delta \phi \sim A\exp\bigl(\pm m x\bigr)$, where the sign depends on the boundary conditions. We can estimate at which value of $\delta\phi$ the leading order approximation breaks down by comparing the size of the next term on the right-hand side. Indeed, the value of $\delta\phi$ for which both terms are equal gives us the rough extent of the tail region, i.e.
\begin{equation}
L_{\rm tail} \equiv \frac{2m^2}{\bigl| V^{\prime\prime\prime}(v)\bigr|}\,.
\end{equation}
The $L_{\rm tail} $ corresponds to the horizontal extend of either of the red regions on Fig.~\ref{fig:potdata}. Note, that the above formula is only valid if $V^{\prime\prime\prime}(v) \not = 0$. If the third derivative vanishes, comparing the leading term with the third order term gives us instead $L_{\rm tail} = \sqrt{6m^2/ \bigl| V^{(4)}(v)\bigr|}$. Similar consideration must be taken when dealing with potentials with vanishing $m$.

The size of the core and skin regions of the potential can be estimated using the same reasoning through expansion around the maximum or an inflection point, respectively. Denoting $\alpha^2 \equiv -V^{\prime\prime}(\phi_{\rm max})$ and $\eta \equiv V^\prime(\phi_{\rm inf})$ the corresponding sizes of these regions read (assuming non-vanishing third derivatives)
\begin{equation}
L_{\rm core} \equiv \frac{2\alpha^2}{\bigl| V^{\prime\prime\prime}(\phi_{\rm max})\bigr|}\,, \quad 
L_{\rm skin} \equiv \sqrt{\frac{2|\eta|}{\bigl|V^{\prime\prime\prime}(\phi_{\rm inf})\bigr|}}\,.
\end{equation}

The numbers $L_{\rm tail}$, $L_{\rm core}$ and $L_{\rm skin}$ should be used as rough gauges of how much the tails, core, and skin dominate the kink's shape. Of course, these regions can have overlaps and typically do not add up to the total value of a field span for a static (anti-)kink, that is $|v_{\rm R}-v_{\rm L}|$.  

For illustration, the corresponding values for a $\phi^4$ potential are $L_{\rm tail}^{\phi^4} = 2/3$, $L_{\rm core}^{\phi^4} = 1$ and $L_{\rm skin}^{\phi^4} = 2/9$, while the same for the sG model reads $L_{\rm tail}^{\rm sG} = \sqrt{6}$,  $L_{\rm core}^{\rm sG} = \sqrt{6}$ and  $L_{\rm skin}^{\rm sG} = \sqrt{2}$. 
The $\phi^6$ potential, having three minima with the middle one of different curvature, thus has two different tails and skin regions, depending on which vacuum and which inflection point is considered. 

Let us also note that there can be additional local minima in between the true vacua that the kink is interpolating. In that case, it is reasonable to dub the corresponding region of the kink near a local minimum as its \emph{pseudocore}. The reason for this nomenclature -- as opposed to calling it a pseudotail -- is simply that a pseudocore region affects the kink's center and it would be confusing to associate it with the semi-infinitely extended tails. In the double sine-Gordon model (dsG), for instance, the kink solution is approximately equal to two sG kinks separated by a fixed distance \cite{Gani:2017yla}. In our picture, each of these sG kinks will have its core and skin regions, while the region between them corresponds to a pseudocore.

\subsection{The Frankensteinian potentials}

So far, we have discussed the structural aspects of a static kink. The issue now is, whether these notions have any relevance for the dynamics and if so, whether their contributions are approximately independent.

Intuitively, the answer should be cautious ``yes''. It is well known, for instance, that kink tails are responsible for attractive interaction between well-separated kink-anti-kink ($K\bar{K}$) pairs and are, therefore key ingredients in the initial (and final) stages of the $K\bar{K}$ collisions \cite{Manton:2004tk}. On the other hand, it is also intuitively clear that they are of little consequence when the kinks are on top of each other where their cores play a key role.\footnote{Assuming, of course, that kinks retain their individuality during the collisions and it is hence still reasonable to talk about their cores. In this regard, the successes of collective coordinate models based on separated kink ansatzes in reproducing qualitative features of $K\bar{K}$ scattering indicate that this is not an unreasonable assumption.}

The role of skin, however, if any, is hard to appreciate intuitively. However, the skin region of the potential (i.e. its inflection point) is deeply connected with the longevity or even existence of oscillons \cite{Gleiser:2008ty}, which are important, if not crucial, for the understanding of the $K\bar{K}$ scattering \cite{Bazeia:2017rxo, Gani:2017yla}.

These observations may help us to appreciate the underlying reason for the complexity of $K\bar{K}$ scattering -- the fact that it may be composed of multiple structural pieces that contribute differently and simultaneously to its dynamics. In other words, it could be the case that $K\bar{K}$ scattering is a tapestry of interwoven but otherwise only loosely dependent contributing phenomena. 

To facilitate this intuition, it should be worthwhile to investigate the dynamics of kinks 
in potentials lacking some or most of these structural pieces. To that end, let us briefly discuss a particular class of potentials that -- for lack of a better term -- we call the \emph{Frankensteinian} potentials. As the name suggests, they are composed of pieces of functions of the field that are continuously and/or differentiably sewn together at chosen field values. For our purposes, we see their primary utility in the fact that inside each patch, a given Frankensteinian potential could be deficient in one or two structural regions. This makes them an ideal tool for exploring the relevance of these regions on the dynamics.

In particular, we will focus on a subset of Frankensteinian potentials that are constructed out of linear or quadratic functions. Within these pieces, the leading order approximations presented above for either a tail, core, or skin hold exactly.
Thusly constructed potentials, therefore, possess the simplest kink solutions as far as their static characteristics are concerned.

\subsection{Single-component kinks}

\begin{figure*}[htb!]
\begin{center}
\includegraphics[width=0.98\textwidth]{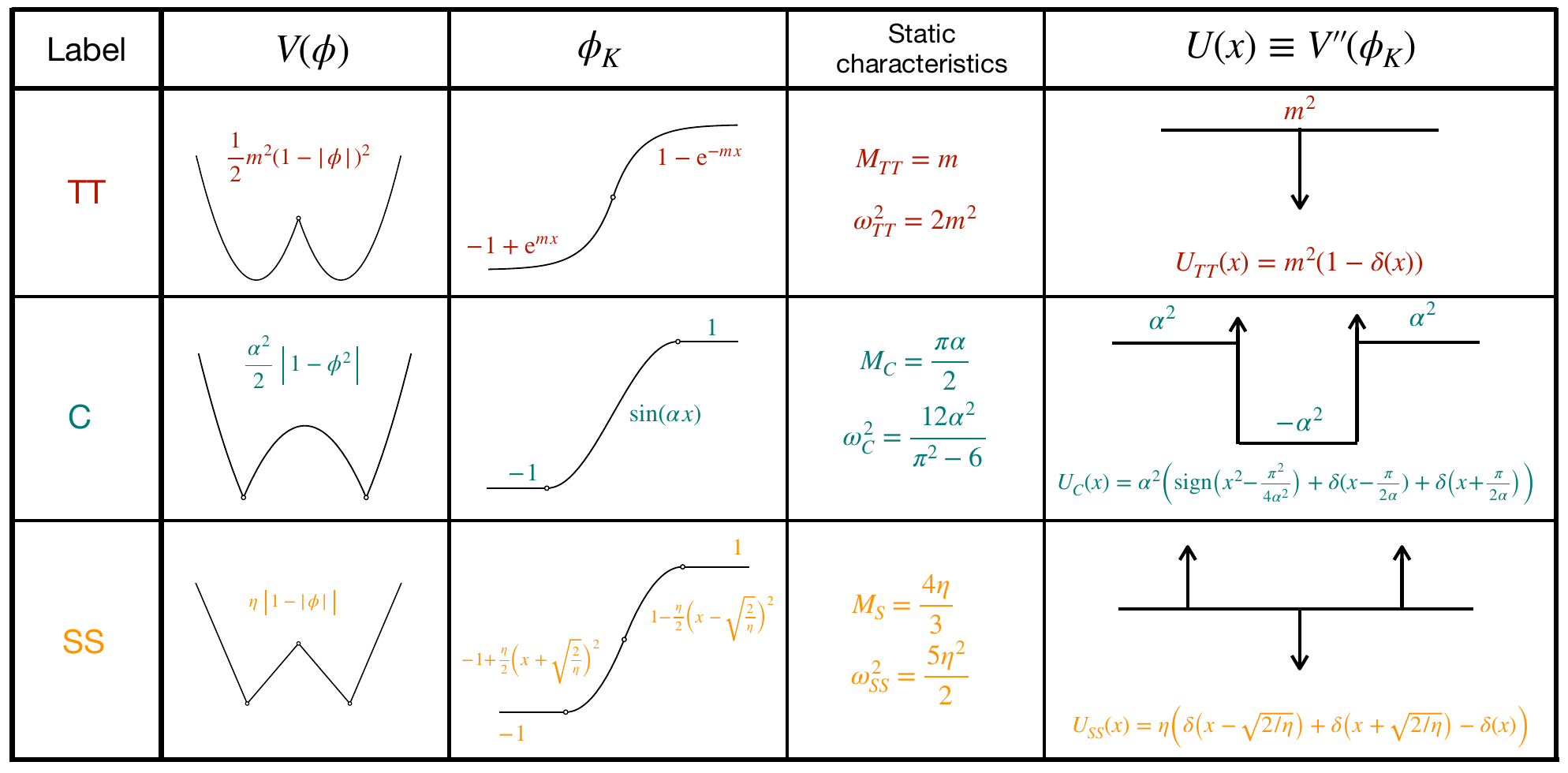}
\end{center}
{
    \caption{\small The ``simplest'' kinks and their potentials. The vacua are placed at $\pm 1$ and kinks are centered at $x=0$ for simplicity. The last row shows the effective potential that enters the Schr\"odinger-like equation for the determination of normal modes of a given kink.}
    \label{fig:simplesttable}
}
\end{figure*}

To provide some examples of Frankensteinian potentials,  let us consider those that support static solutions of the greatest structural simplicity, namely kinks made of a single type of component. In Fig.~\ref{fig:simplesttable}, we present basic characteristics for these potentials and their kinks. We use the labeling $TT$, $C$, and $SS$ that simply tally the structural pieces of the kinks as seen from the left to right.

The potential labeled by $TT$ (``tail-tail'') consists of two quadratic wells sewn together at the center $\phi=0$. For simplicity, both wells have the same curvature $m^2$, and the respective vacua are placed at $\pm1$, which is always possible to enforce via rescaling of the field. In fact, it is also possible to set $m=1$ by rescaling the coordinates. 

Consequently, the $TT$ kink is made of two exponential tails sewn differentiably together at the center of the kink where $\phi =0$. With the energy density given as 
\begin{equation}
\mathcal{E}_{TT} = m^2 \Exp{-2 m |x|}\,,
\end{equation}
the mass of the $TT$ kink works out to be equal to the perturbative mass, i.e. $M_{TT} \equiv \int\limits_{-\infty}^{\infty} \mathcal{E}_{TT}\, \diff x = m$. In turn, the second moment of energy density reads $\int\limits_{-\infty}^{\infty}x^2 \mathcal{E}_{TT}\, \diff x = 1/(2m)$, which provides a simple measure of how the energy is concentrated around the kink's center. 

This can be equivalently expressed using the so-called Derrick's frequency, which is defined as the ratio of the mass and the second moment and for a $TT$ kink reads $\omega_{TT}^2 = 2 m^2$. 

Derrick's frequency is associated with Derrick's mode of the kink, which is derived by observing how the energy of a static kink changes under an infinitesimal scaling of the spatial coordinate. Derrick's mode, however, is not a normal mode, but it has been recognized to be important for restoring Lorentz invariance of collective coordinate models  \cite{Adam:2021gat}. That being said, the $TT$ kink has Derrick's frequency well above the mass threshold and hence should have no impact on the dynamics.

The normal modes of $TT$ kink can be also easily established by solving the Schr\"odinger-like equation with the effective potential given by $U(x) \equiv V^{\prime\prime}(\phi_{TT})$. As we see in the last row of Fig.~\ref{fig:simplesttable}, the effective potential for $TT$ kink consists of a single $\delta$ well with strength $-m^2$ plus a constant $m^2$. As is well known, a Dirac $\delta$-well potential supports only a single bound mode, which is the zero mode associated with uniform translations. Hence, no massive bound modes exist for $TT$ kink. Note, that the effective potential $U(x)$ is the only quantity in Fig.~\ref{fig:simplesttable} that is sensitive to the regions of the potential outside the vacua, i.e. $|\phi|>1$.  The kink solution itself and its static characteristics given in the fourth row of Fig.~\ref{fig:simplesttable} would remain unaltered if the potential is modified in these outer regions.

Let us now briefly turn our attention to the remaining two kinks, that are made of a core ($C$) and two skins $(SS)$. These represent the simplest examples of  \emph{compact} kinks whose spatial extents are finite. This is due to the non-analytic minima, which effectively means an infinite perturbative mass, rendering the tails non-existent. We will not comment on the dynamics of these kinks in this paper outside what is given in Fig.~\ref{fig:simplesttable}. 
The reason is that these compact kinks have been proposed and investigated before. 
In fact, the origin of compact solitons can be traced back to the 90ties, when they were introduced in the context of modified KdV equation \cite{Rosenau:1993zz}, investigation of which continues to this day \cite{Francisco}. 

The compact solitons in the relativistic scalar field theory were first explored in the early 2000s in \cite{Arodz:2002yt}, later in the context of the so-called $V$-shaped potentials \cite{Arodz:2005gz}, the most famous of which is the so-called signum-Gordon potential \cite{Arodz:2007ek}. The signum-Gordon model is probably the simplest Frankensteinian model consisting only of two linear functions sewn at the middle forming a $V$-shaped potential well, which supports exactly soluble compact oscillons \cite{Arodz:2007jh, Hahne:2019ela}. Furthermore, the compact kinks were introduced as a limit of certain mechanical linear systems with chained pendula in \cite{Arodz, Arodz:2003mx}. 

The Frankensteinian potential that was proposed for studying the interaction between compact kinks and oscillons \cite{Hahne:2022wyl} and the scattering of compact kinks themselves \cite{Hahne:2023dic} was a periodic version of the potential $C$ in Fig.~\ref{fig:simplesttable}.
As shown in \cite{Hahne:2023dic}, collisions of compact $K\bar{K}$ pairs lead to long-living oscillating bound states or the $K\bar{K}$ pair reemerges with accompanying ``shockwave'' that disintegrates into a cascade of compact oscillons. Curiously, the characteristic bouncing of kink anti-kink has not been observed. This is, perhaps, because the potential used in \cite{Hahne:2023dic} was a periodic piece-wise quadratic function that lacked any other non-linearities besides the non-analytic sewing at minima. 

As far as we are aware, no dedicated study has been published on compact kinks made of skins.

A different kind of Frankensteinian potential was also presented in \cite{campostoymodel}, which is a piece-wise quadratic potential resembling $\phi^4$ potential with the sewing points around the region controlling the skin of the kink. As a result, the kink solution possesses both exponentially dampened tails and a core region, but no skin. In our notation, such a solution is a symmetric $TCT$ kink. These kinks feature all characteristics of generic $K\bar{K}$ scattering: bouncing, bion formation, and radiation production.

Let us also mention that there have been few works that investigate potentials near Frankensteinian (or otherwise singular) limits that support compact kinks \cite{Bazeia:2014hja, Bazeia:2019tgt}. 

The central task of this paper is to both continue the investigation of the Frankensteinian potentials and to address a specific question: namely how does the core region affect the outcomes of $K\bar{K}$ scattering? In particular, we will focus on the presence or absence of characteristic dynamical feature due to the resonant-energy transfer mechanism, namely the bouncing.

In Sec.~\ref{sec:II}, we begin addressing this question by briefly discussing $K\bar{K}$ scattering in models of increasing structural complexity. We will first show that scattering of $TT$ kinks is particularly simple: the only outcome is a total annihilation into massive waves. Then we show that this result is unaltered by adding non-linearities inside the tail region, without introducing inflection points. It is only in the third example, with a potential possessing inflection points and therefore skin region, where we see oscillons together with critical velocity below one.

In Sec.~\ref{sec:III} we change tactics and present two parametric families of potentials interpolating between $\phi^4$ and particular coreless potentials. Here we study the disappearance of resonant structures in $K\bar{K}$ collisions as the potential becomes more and more singular at its center.

In Sec.~\ref{sec:V} we discuss our results.

%%%%%%%%%%%%%%%%%%%%%%%%%%%%%%%%%%%%%%%%%%%%%%%%%%%%%%%%%%
%%%%%%%%%%%%%%%%%%%%%%%%%%%%%%%%%%%%%%%%%%%%%%%%%%%%%%%%%%
%%%%%%%%%%%%%%%%%%%%%%%%%%%%%%%%%%%%%%%%%%%%%%%%%%%%%%%%%%
%%%%%%%%%%%%%%%%%%%%%%%%%%%%%%%%%%%%%%%%%%%%%%%%%%%%%%%%%%
\section{Coreless kinks: three examples}
\label{sec:II}
%%%%%%%%%%%%%%%%%%%%%%%%%%%%%%%%%%%%%%%%%%%%%%%%%%%%%%%%%%
%%%%%%%%%%%%%%%%%%%%%%%%%%%%%%%%%%%%%%%%%%%%%%%%%%%%%%%%%%
%%%%%%%%%%%%%%%%%%%%%%%%%%%%%%%%%%%%%%%%%%%%%%%%%%%%%%%%%%
%%%%%%%%%%%%%%%%%%%%%%%%%%%%%%%%%%%%%%%%%%%%%%%%%%%%%%%%%%

In this section, we shall consider three examples of coreless potentials. As we will see, for the first two, the absence of both core and skin regions seems to render the $K\bar{K}$ scattering entirely trivial: the only result of the collision is annihilation into massive waves. Only in the third example, where we consider a potential with inflection point = non-zero skin region, do we obtain oscillons, however, the bouncing is still absent.

\subsection{Scattering of $TT$ kinks}

Let us comment on $K\bar{K}$ scattering of the kinks made from only exponential tails. As already described, the potential is
constructed by gluing together two quadratic wells of the same curvature at the origin. We can express such potential as
 \begin{align}\label{eq:cusp}
V_{TT}(\phi) = & \frac{1}{2}m^2 \bigl(1-|\phi|\bigr)^2\,.
\end{align}
Compared with $\phi^4$ potential, the TT potential lacks analytic maximum, inflection points and has no nonlinearities besides the sewing point at $\phi=0$. This makes the dynamics of TT kinks predictably featureless as we shall see.

\begin{figure}[htb!]
\begin{center}
\includegraphics[width=0.7\columnwidth]{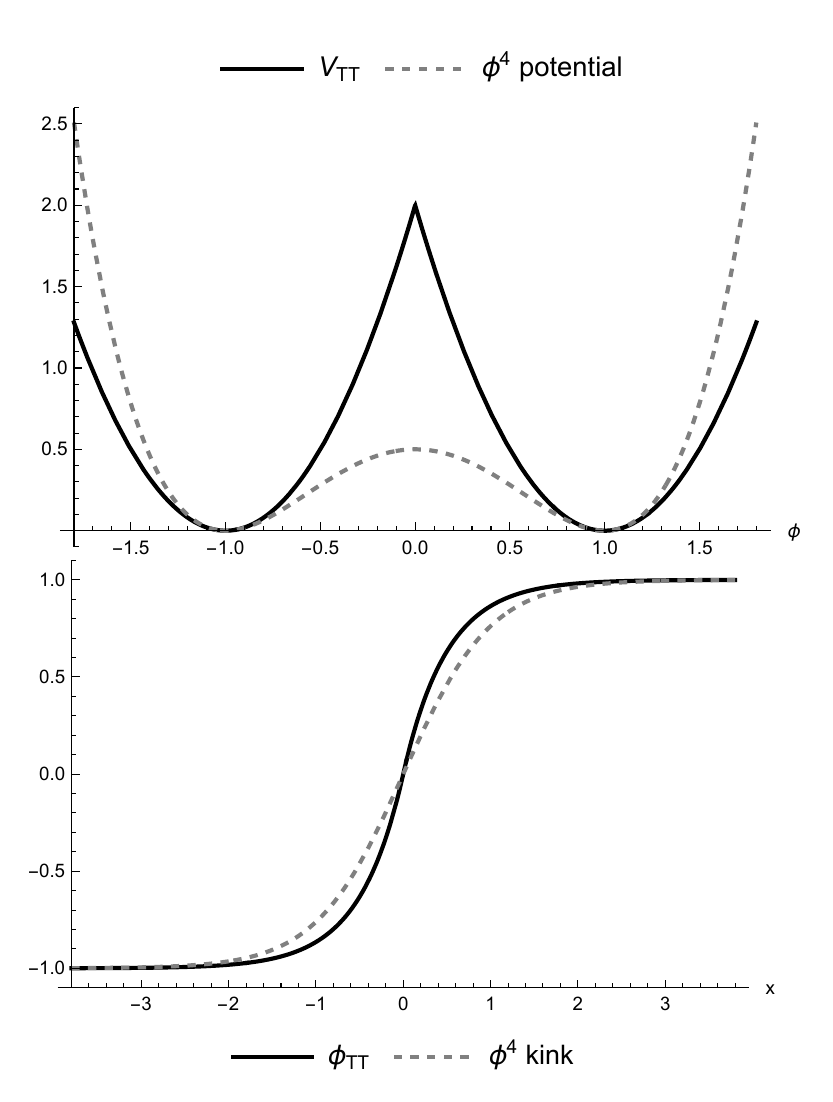}
\end{center}
{
    \caption{\small An example of piece-wise quadratic potential with the $\myphi$ kink solution compared with $\phi^4$ model.}
    \label{fig:pot}
}
\end{figure}

The kink solution can be obtained by sewing together two exponentials. Up to arbitrary shift along the $x$-axis, the solution reads:
\begin{gather} \label{eq:coreless2}
\myphi = {\rm sign}(x)\Bigl(1- \Exp{-m |x|}\Bigr)\,,
\end{gather}
where ${\rm sign}(x)$ is the sign function. Both the potential \refer{eq:cusp} and the kink solution are illustrated in Fig.~\ref{fig:pot}.

It is easy to verify that $\myphi$ solves the second-order equations of motion in a weak sense. There are no $\delta$-function contributions since not only the field itself but also its first derivative is continuous. However, there is a jump singularity for the second derivative.

As discussed in the previous section, there are no massive normal modes and Derrick's frequency is above the mass threshold, i.e. $\omega_D^2 = 2m^2 > m^2$.
It is thus not surprising that collisions of a TT kink with an anti-kink are quite boring. However, this featureless-ness is of a different kind than, say in sG model, where $K\bar{K}$ pairs collide elastically and do not annihilate each other due to the underlying integrability. In contrast, $TT$ kinks \emph{always annihilates} each other. 

It is easy to understand why. When the $K\bar{K}$ pair is sufficiently close to each other, the whole field becomes localized entirely within the left quadratic well. From that point on, the dynamics is equivalent to a time-evolution of some initial data via the Klein-Gordon equation with the mass $m$. As is well known, this results in the disintegration of the initial shape into a train of massive waves.

\begin{figure}[htb!]
\begin{center}
\includegraphics[width=0.99\columnwidth]{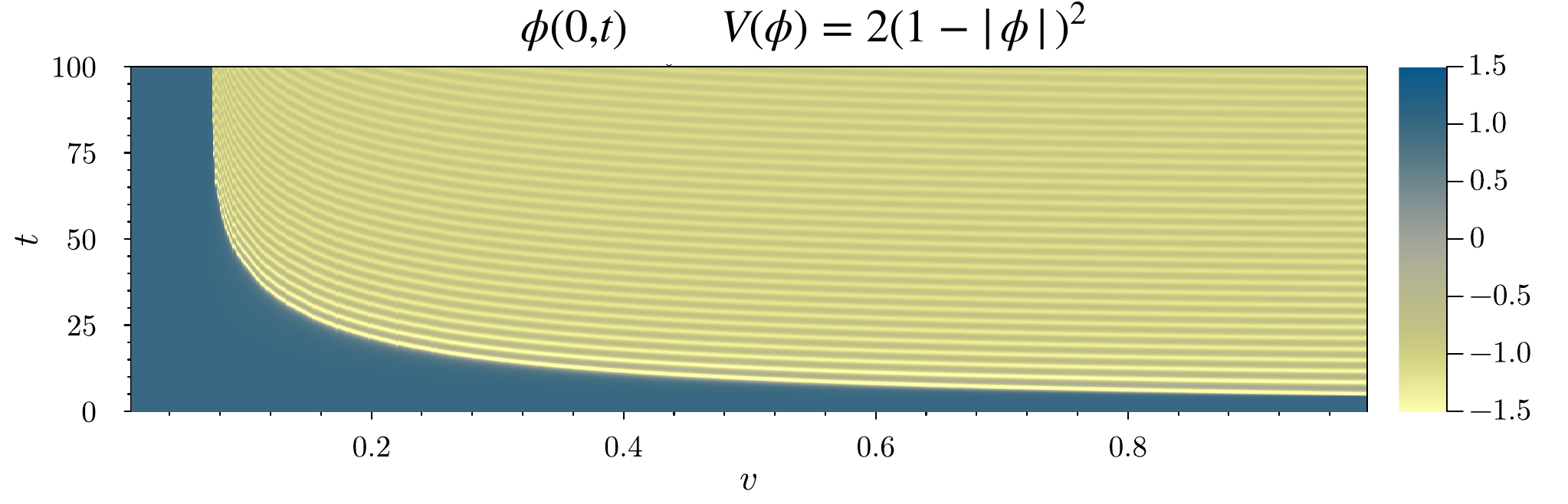}
\end{center}
{
    \caption{\small A plot of values of $\phi(0,t)$ as dependent on initial velocity $v$ of the $K\bar{K}$ pair. We see that for any $v$, the initially separated pair (blue color indicating $\phi\sim +1$ vacuum at the center before the collision) always disintegrates into massive waves around the $\phi\sim -1$ vacuum (the yellow colors). Here, $m=2$.}
    \label{fig:TTcollisionmap}
}
\end{figure}

As a check of the above intuition, we plot a ``map'' of $K\bar{K}$ scattering in Fig.~\ref{fig:TTcollisionmap}, where we show the dependence of the central value of the field on time, i.e. $\phi(0,t)$, for the whole range of initial velocities of the $K\bar{K}$ pair.

\subsection{Scattering of kinks in piece-wise quartic wells.}

The triviality of scattering of $TT$ kinks can be blamed on the linearity of the respective equation of motion in each half-plane of the target space. 
Therefore, it is unclear whether this dull result has anything to do with the absence of core and skin region in the $V_{TT}(\phi)$ potential or, rather, whether it is an artifact of its piece-wise integrability. For this reason, in this subsection, we will investigate a potential that is non-linear in each potential well. This is achieved by adding a quartic term, i.e.
\begin{equation}\label{eq:TTlambdapot}
V_{TT\lambda}(\phi) \equiv \frac{1}{2}m^2 \bigl(1-|\phi|\bigr)^2+ \frac{\lambda}{12} \bigl(1-|\phi|\bigr)^4\,,
\end{equation}
where $\lambda >0$ is an arbitrary positive constant.
This potential leads to the equation of motion that is neither integrable nor linear in the respective halves of the target space.  However, it still lacks inflection points, and thus the kink solution, i.e.
{\small \begin{equation}
\phi_{TT\lambda}= {\rm sign}(x)\biggl(1-\frac{m\sqrt{6/\lambda}}{\sinh\bigl(m |x|+\sinh^{-1}(m\sqrt{6/\lambda})\bigr)}\biggr)\,,
\end{equation}}
has neither a core nor a skin, but possesses non-linear tails.

\begin{figure}[htb!]
\begin{center}
\includegraphics[width=0.7\columnwidth]{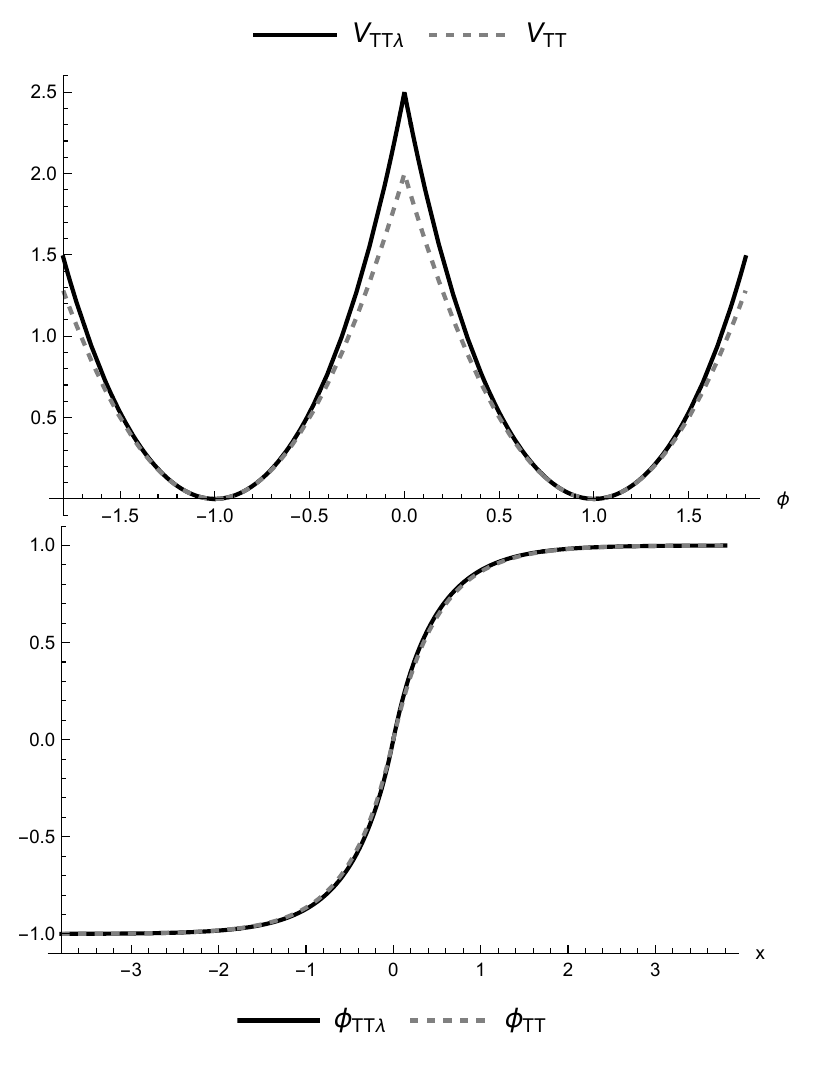}
\end{center}
{
    \caption{\small The potential \refer{eq:TTlambdapot} and its kink solution compared with $\lambda=0$ case. Here $m=2$ and $\lambda = 6$.}
    \label{fig:TTlambdapot}
}
\end{figure}

Notice that in the limit $\lambda \to 0$, the solution becomes ${\rm sign}(x)\bigl(1-\Exp{-m|x|}\bigr)$, which is nothing but a $TT$ kink. We display both the potential and its kink solution in Fig.~\ref{fig:TTlambdapot}.

The Bogomol'nyi-Prasad-Sommerfield (BPS) mass \cite{Manton:1981mp, Bogomolny:1975de} of the kink reads
\begin{equation}
M_{TT\lambda} = \frac{4 m^3}{\lambda}\biggl(\Bigl(1+\frac{\lambda}{6m^2}\Bigr)^{3/2}-1\biggr)\,,
\end{equation}
which is an increasing function of $\lambda$. Derrick's frequency can be also shown to be an increasing function of $\lambda$. Hence, it always stays above the perturbative threshold for all $\lambda>0$. We also did not find any massive modes. Indeed, a quick glance at the effective potential $V^{\prime\prime}(\phi_{TT\lambda})$ reveals that the extra term acts as a potential barrier rather than a well, hence there is no hope for any bound mode besides the zero mode that is there due to the delta-peak potential well at $x=0$. 

\begin{figure}[htb!]
\begin{center}
\includegraphics[width=0.99\columnwidth]{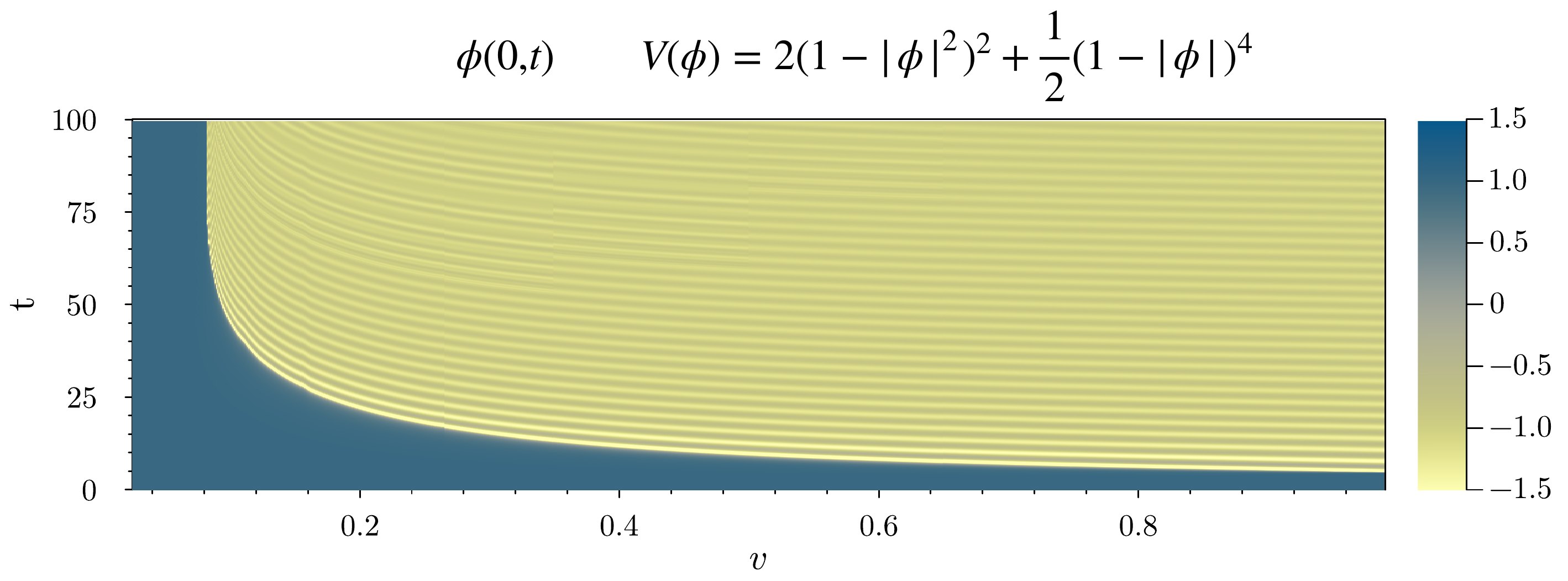}
\end{center}
{
    \caption{\small The plot of values of $\phi(0,t)$ as dependent on the initial velocity $v$ of the $K\bar{K}$ pair in the model with additional quartic term.}
    \label{fig:TTlambdacollisionmap}
}
\end{figure}

Given that there are no modes that could facilitate resonant energy transfer, we expect that the dynamics of $K\bar{K}$ scattering is equally trivial as for $TT$ kinks. Indeed, this is what we found. We show our result in Fig.~\ref{fig:TTlambdacollisionmap}.

\subsection{Scattering of kinks in log-corrected Klein-Gordon model}

\begin{figure}[htb!]
\begin{center}
\includegraphics[width=0.7\columnwidth]{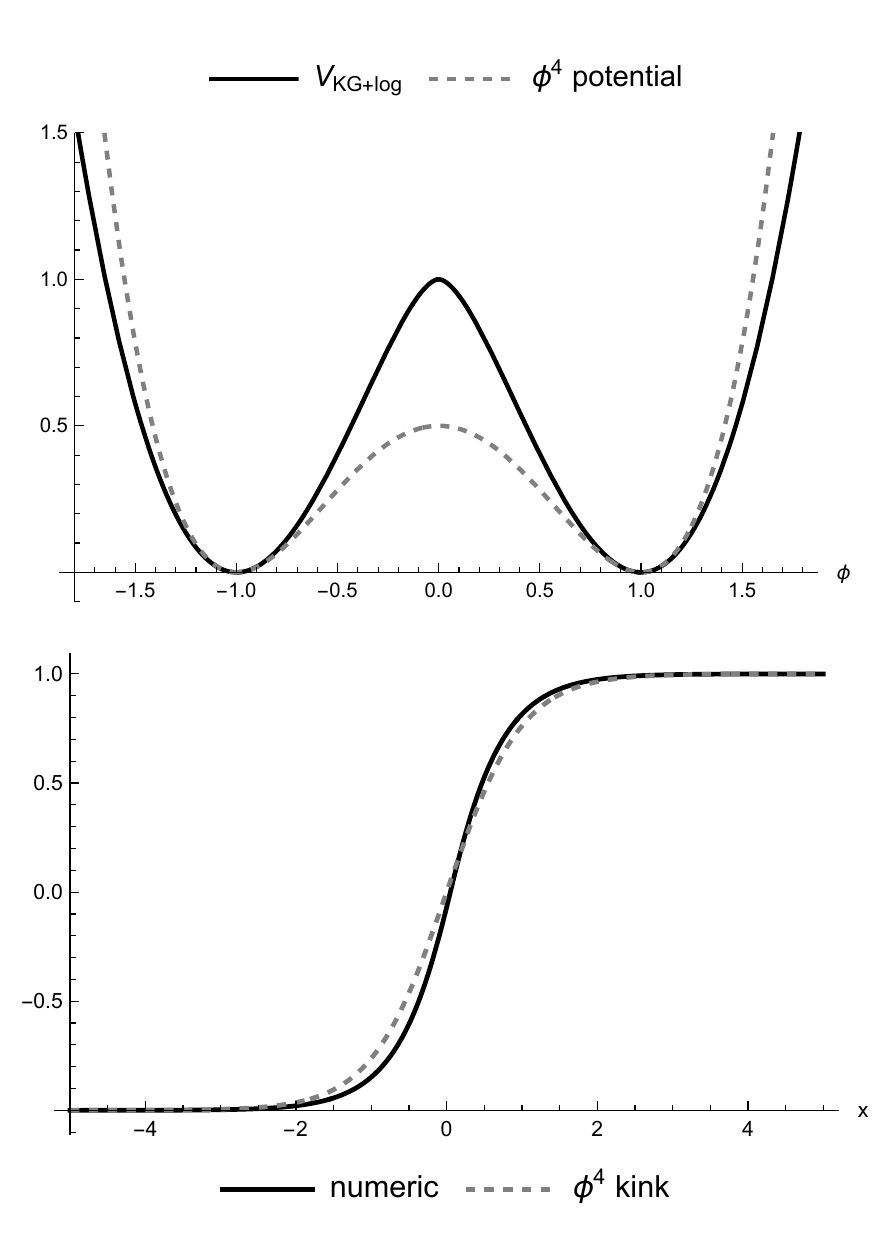}
\end{center}
{
    \caption{\small The potential \refer{eq:KG+log} and its kink solution compared with $\phi^4$ model. Here $m=2$.}
    \label{fig:KG+log}
}
\end{figure}

Lastly, let us consider a coreless potential, which does have inflection points, namely the Klein-Gordon model with a logarithmic ``correction'', i.e.
\begin{equation}\label{eq:KG+log}
V_{\rm KG+log}(\phi) = \frac{1}{4}m^2 \Bigl(1-\phi^2+\phi^2 \log(\phi^2)\Bigr)\,.
\end{equation} We display it in Fig.~\ref{fig:KG+log}.

 This potential has a degenerate minima at $\phi = \pm 1$ with $V^{\prime\prime}(\pm1) = m^2$. Note that we can fix $m^2$ to any value by rescaling the coordinates. As was the case for the previous two potentials, the maximum at $\phi=0$ is non-analytic since the second derivative at $\phi=0$ is undefined (it goes to $-\infty$). This makes the Taylor expansion around the origin impossible, rendering $V_{\rm KG+log}(\phi)$ coreless.
However, unlike for both $V_{TT}(\phi)$ and $V_{TT\lambda}(\phi)$, there are inflection points located at $\phi_{\rm inf} = \pm 1/{\rm e}$, hence it makes sense to talk about the skin.

The kink solution cannot be obtained analytically and we showcase a numerical solution in Fig.~\ref{fig:KG+log}. The BPS mass reads
{\small \begin{equation}
M_{\rm KG+log} = \frac{m}{\sqrt{2}}\int\limits_{-1}^{1}\sqrt{\eta^2 \log \eta^2 +1-\eta^2}\, \diff \eta \approx 0.84 \times m\,.
\end{equation}}

Derrick's frequency is numerically determined to be $\omega_D^2 \approx 1.34 m^2$, so it is above the perturbative threshold. Similarly, we have not found any massive mode. 
Given that there are no bound modes, we could expect the scattering of $K\bar{K}$ pairs to be still completely sterile as in the previous two models, but this expectation would be incorrect. 

\begin{figure*}[htb!]
\begin{center}
\includegraphics[width=0.9\textwidth]{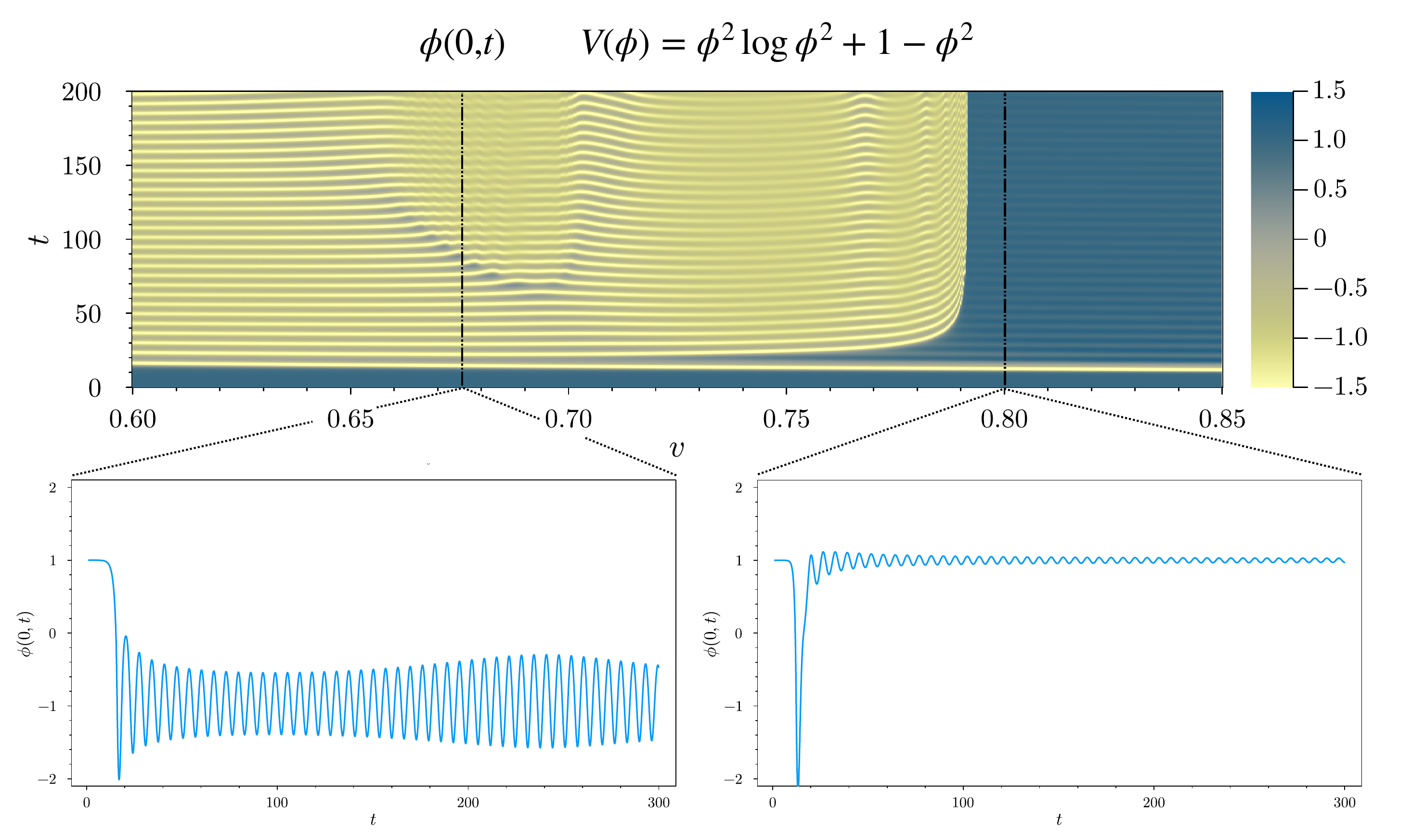}
\end{center}
{
    \caption{\small The evolution of the central value of the field $\phi(0,t)$ as a function of initial velocity of the $K\bar{K}$ pair. In the insets, we see two particular choices of $v$ that showcase the formation of an oscillon and a quasi-elastic collision. Here $m=2$.}
    \label{fig:KG+logcollisions}
}
\end{figure*}

In fact, compared with $V_{TT}(\phi)$ and $V_{TT\lambda}(\phi)$, the $K\bar{K}$ scattering in $V_{\rm KG+log}$ potential exhibits two new features. The first one is the appearance of a critical velocity, $v_{\rm crit} \approx 0.79$, above which the scattering is quasi-elastic, and second, there are certain velocities below $v_{\rm crit}$ for which oscillons are produced. This can be seen in Fig.~\ref{fig:KG+logcollisions}.\footnote{This potential has been studied in greater depth in \cite{Belendryasova:2021jgs}. The findings presented in this subsection have been obtained independently and agree with those presented in \cite{Belendryasova:2021jgs}.}

The results of this section show that if a potential lacks both core and skin regions, it implies a lack of any interesting dynamics in $K\bar K$ collisions. This was true for both $V_{TT}(\phi)$ and $V_{TT\lambda}(\phi)$. When we considered coreless potential with inflection points,  i.e. $V_{\rm KG+log}(\phi)$, we immediately got more interesting dynamics. Of course, we cannot prove that this must be the case in general, but we take these observations as a confirmation that the core and skin regions greatly affect the outcomes of $K\bar{K}$ collisions.

%%%%%%%%%%%%%%%%%%%%%%%%%%%%%%%%%%%%%%%%%%%%%%%%%%%%%%%%%%
%%%%%%%%%%%%%%%%%%%%%%%%%%%%%%%%%%%%%%%%%%%%%%%%%%%%%%%%%%
%%%%%%%%%%%%%%%%%%%%%%%%%%%%%%%%%%%%%%%%%%%%%%%%%%%%%%%%%%
%%%%%%%%%%%%%%%%%%%%%%%%%%%%%%%%%%%%%%%%%%%%%%%%%%%%%%%%%%
\section{Disappearance of bouncing windows and non-analyticity of the maximum}
\label{sec:III}
%%%%%%%%%%%%%%%%%%%%%%%%%%%%%%%%%%%%%%%%%%%%%%%%%%%%%%%%%%
%%%%%%%%%%%%%%%%%%%%%%%%%%%%%%%%%%%%%%%%%%%%%%%%%%%%%%%%%%
%%%%%%%%%%%%%%%%%%%%%%%%%%%%%%%%%%%%%%%%%%%%%%%%%%%%%%%%%%
%%%%%%%%%%%%%%%%%%%%%%%%%%%%%%%%%%%%%%%%%%%%%%%%%%%%%%%%%%

In this section, we investigate $K\bar{K}$ scattering in two parametric families of models that start from $\phi^4$ potential and develop a cusp singularity around the origin ($ V(\phi) \sim |\phi|$) as $\varepsilon$ goes from 1 to 0. 
In this way, we explore how the rate of divergence of $V^{\prime\prime}(\phi)$ at $\phi\to 0$ affects the bouncing windows. As we shall see, the bouncing windows disappear in an almost linear fashion well before the point $\varepsilon=0$ is reached.

\subsection{Two families of coreless potentials}

\begin{figure}[htb!]
\begin{center}
\includegraphics[width=0.90\columnwidth]{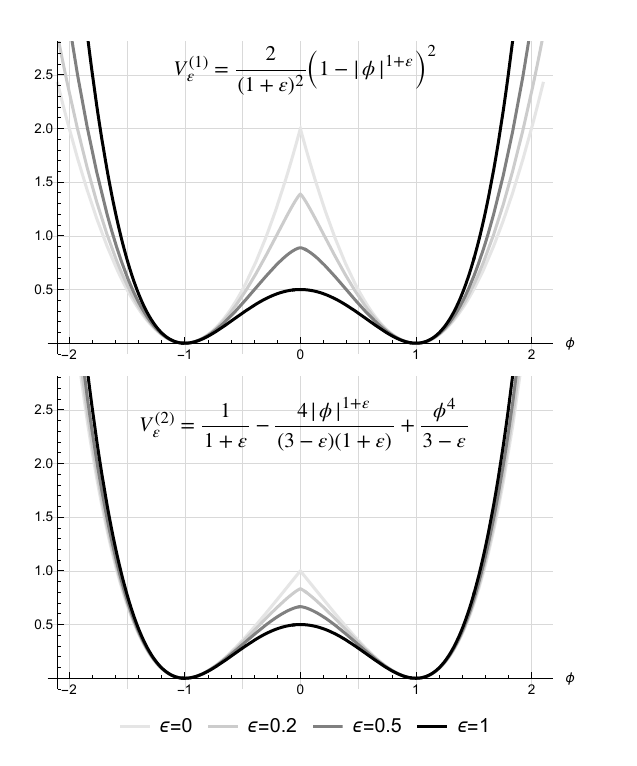}
\end{center}
{
    \caption{\small Two families of potentials that start as $\phi^4$ model (thick black line)  and gradually develop a cusp singularity at the central hill as $\varepsilon \to 0$.}
    \label{fig:potentials}
}
\end{figure}

We define the first family in such a way that it begins as $\phi^4$ potential at $\varepsilon=1$  and deforms into $V_{TT}$ potential \refer{eq:cusp} as $\varepsilon \to 0$, i.e.
\begin{equation}\label{eq:family}
V_{\varepsilon}^{(1)}(\phi) = \frac{2}{(1+\varepsilon)^2}\Bigl(1-|\phi|^{1+\varepsilon}\Bigr)^2\,.
\end{equation} 
We display $V_{\varepsilon}$ in Fig.~\ref{fig:potentials} (top).

The second family aims to deform the $\phi^4$ model only around the maximum and keep the quartic term for all $\varepsilon \leq 1$. 
\begin{equation}\label{eq:famil2}
V_{\varepsilon}^{(2)}(\phi) =\frac{1}{1+\varepsilon}-\frac{4 |\phi|^{1+\varepsilon}}{(3-\varepsilon)(1+\varepsilon)} +\frac{\phi^4}{3-\varepsilon}\,.
\end{equation} 
We display $V_{\varepsilon}$ in Fig.~\ref{fig:potentials} (bottom).

Note that both families are coreless in the entire range $\varepsilon\in [0,1)$ as $V^{\prime\prime}(0)$ is undefined.

As $\varepsilon \to 0$, the potential $V_{\varepsilon}^{(1)}$ looses its skin regions. Indeed, 
the positions of inflection points, given as $|\phi_{\rm inf}^{(1)}| = (\varepsilon/(1+2\varepsilon))^{1/(1+\varepsilon)}$, gradually decreases from $\phi^4$ values $\pm 1/\sqrt{3}$ to zero as $\varepsilon \to 0$, so that they merge with the maximum.
However, in $V_{\varepsilon}^{(2)}$ family, the inflection points stay separate from the central maximum for all $\varepsilon$. Indeed, $|\phi_{\rm inf}^{(2)}| = \bigl(1/3\bigr)^{1/(3-\varepsilon)}$ and the $\varepsilon=0$ positions are $\pm 1/\sqrt[3]{3}$.

For both $V_{\varepsilon}^{(1)}$ and $V_{\varepsilon}^{(2)}$ the curvatures of the vacua remain constant, i.e. $m^2 = 4$. 

\begin{figure}[htb!]
\begin{center}
\includegraphics[width=0.90\columnwidth]{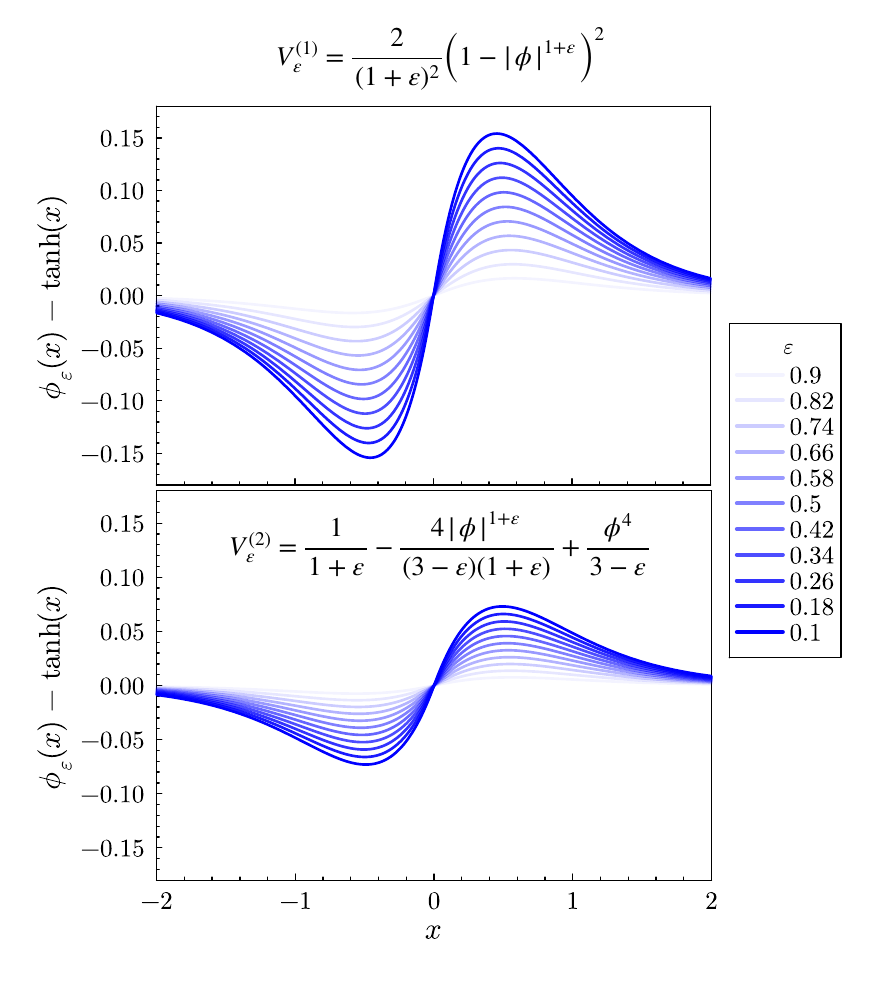}
\end{center}
{
    \caption{\small Absolute differences between shapes of deformed kinks and $\phi^4$ kink for various values of  $\varepsilon $ for both families given in Eqs.~\refer{eq:family}-\refer{eq:famil2}.}
    \label{fig:kinks}
}
\end{figure}

For the first family, the kink solution can be given implicitly as
\begin{equation}
\phi\ _2 F_1 \biggl(\begin{matrix}1\quad 1/(1+\varepsilon) \\  (2+\varepsilon)/(1+\varepsilon) \end{matrix}\biggr| \ |\phi|^{1+\varepsilon}\biggr) = \frac{2 x}{1+\varepsilon}\,,
\end{equation}
where $_2 F_1$ is the hypergeometric function. For the second family, the BPS equation can be solved only numerically. However, for both families,  the shapes of kinks remain quite close to the  $\phi^4$ kink, i.e.  $\tanh(x)$, as seen from Fig.~\ref{fig:kinks}. As expected, the absolute deviation is smaller for $V_{\varepsilon}^{(2)}$.

\begin{figure}[htb!]
\begin{center}
\includegraphics[width=0.90\columnwidth]{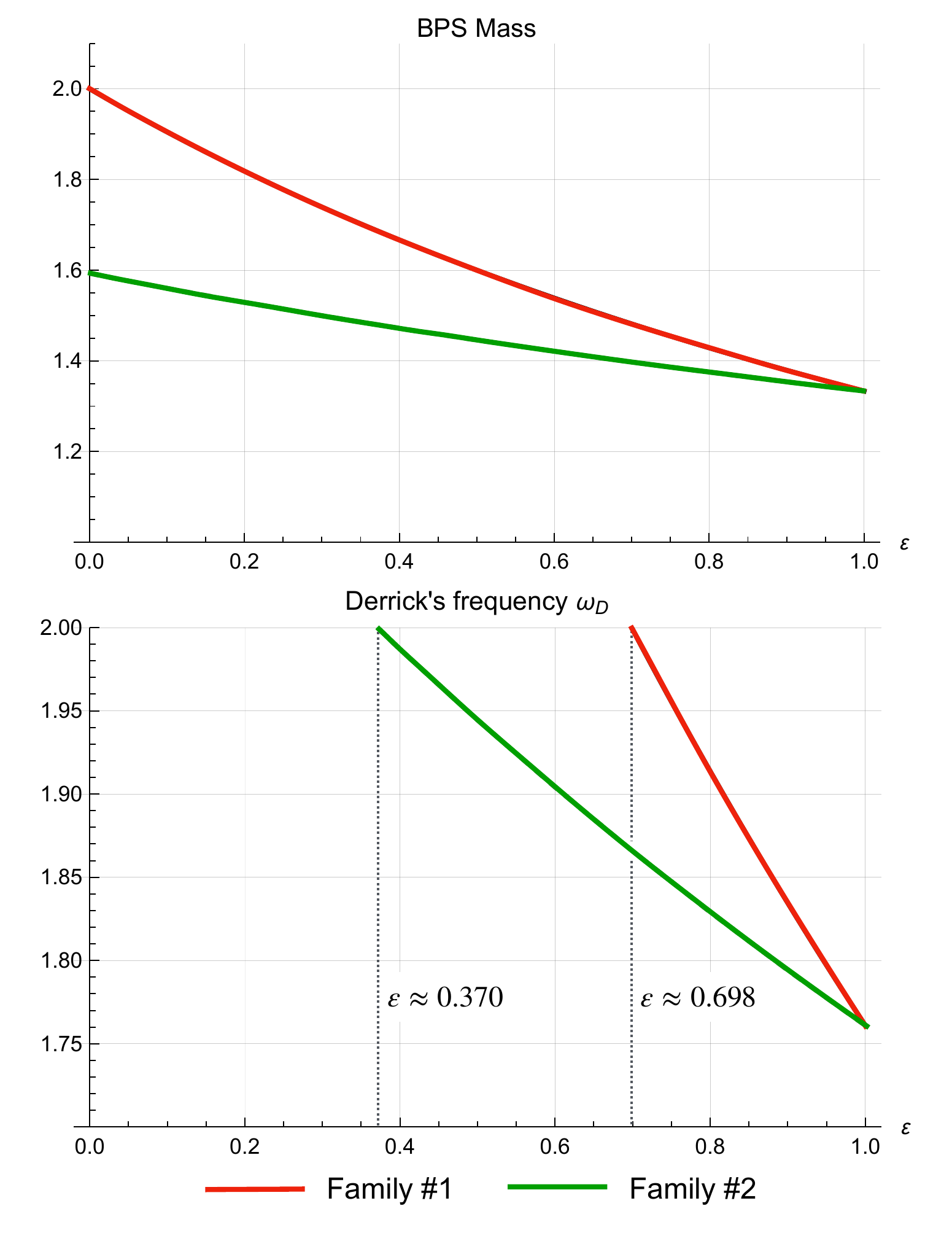}
\end{center}
{
    \caption{\small The BPS masses and Derrick's frequencies for families as functions of $\varepsilon$.}
    \label{fig:MassDerrick}
}
\end{figure}

The BPS mass of the first family can be obtained analytically as $M_K^{(1)} = 4/(2+\varepsilon)$, while for the second family it is known only implicitly through the integral $2\int\limits_{0}^{1} \diff\phi\, \sqrt{2V_{\varepsilon}^{(2)}}$. The frequencies of Derrick's modes must be obtained numerically for both families. We display both these quantities in Fig.~\ref{fig:MassDerrick}.

\begin{figure}[htb!]
\begin{center}
\includegraphics[width=0.90\columnwidth]{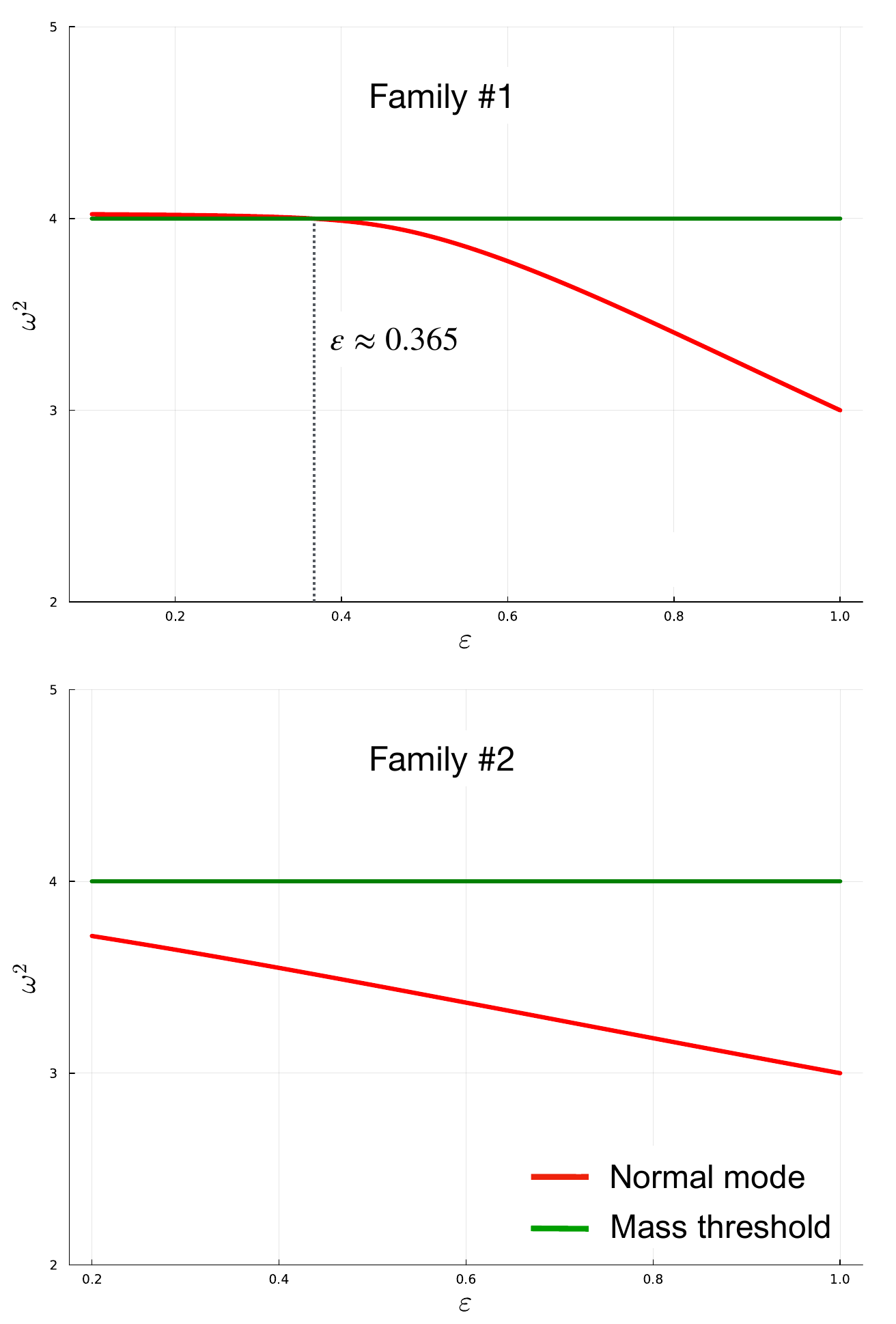}
\end{center}
{
    \caption{\small Dependence of normal modes on $\varepsilon$ for families $V_{\varepsilon}^{(1,2)}$. For  $V_{\varepsilon}^{(1)}$, the massive mode crosses the continuum threshold around $\varepsilon_{\rm threshold} = 0.365$. 
 }
    \label{fig:modesmodels12}
}
\end{figure}

An interesting question is what happens to the structure of normal modes as $\varepsilon \to 0$? In particular, we are interested in the change of the sole bound mode of $\phi^4$ kink with the frequency $\omega = \sqrt{3}$. In Fig.~\ref{fig:modesmodels12} we see that with decreasing $\varepsilon$, the frequency of this mode gradually increases and merges with the continuum around the value $\varepsilon_{\rm threshold} \approx 0.365$. On the other hand, for the second family, the massive mode never merges with the continuum.

\subsection{Bouncing windows}

Given how we set up our models, it is expected that all dynamical features connected with resonant energy transfer disappear for the first family $V_{\varepsilon}^{(1)}$ as $\varepsilon \to 0$. The only question is how fast.

We explore this issue by performing numerical analysis of $K\bar{K}$ collisions (see details of our numerical method in the Appendix) for various values of $\varepsilon$. The main result is displayed in Fig.~\ref{fig:maps} (left column). Surprisingly, the bouncing windows disappear well before $\varepsilon_{\rm threshold}$ where the massive mode crosses into the continuum. In fact, the actual value $\varepsilon \approx 0.7$ coincides almost perfectly with the merging of Derrick's frequency into the continuum (see Fig.~\ref{fig:MassDerrick}). This is surprising, as it is usually the massive mode and not Derrick's mode that is linked with the presence of the bouncing windows.

\begin{figure}[htb!]
\begin{center}
\includegraphics[width=0.90\columnwidth]{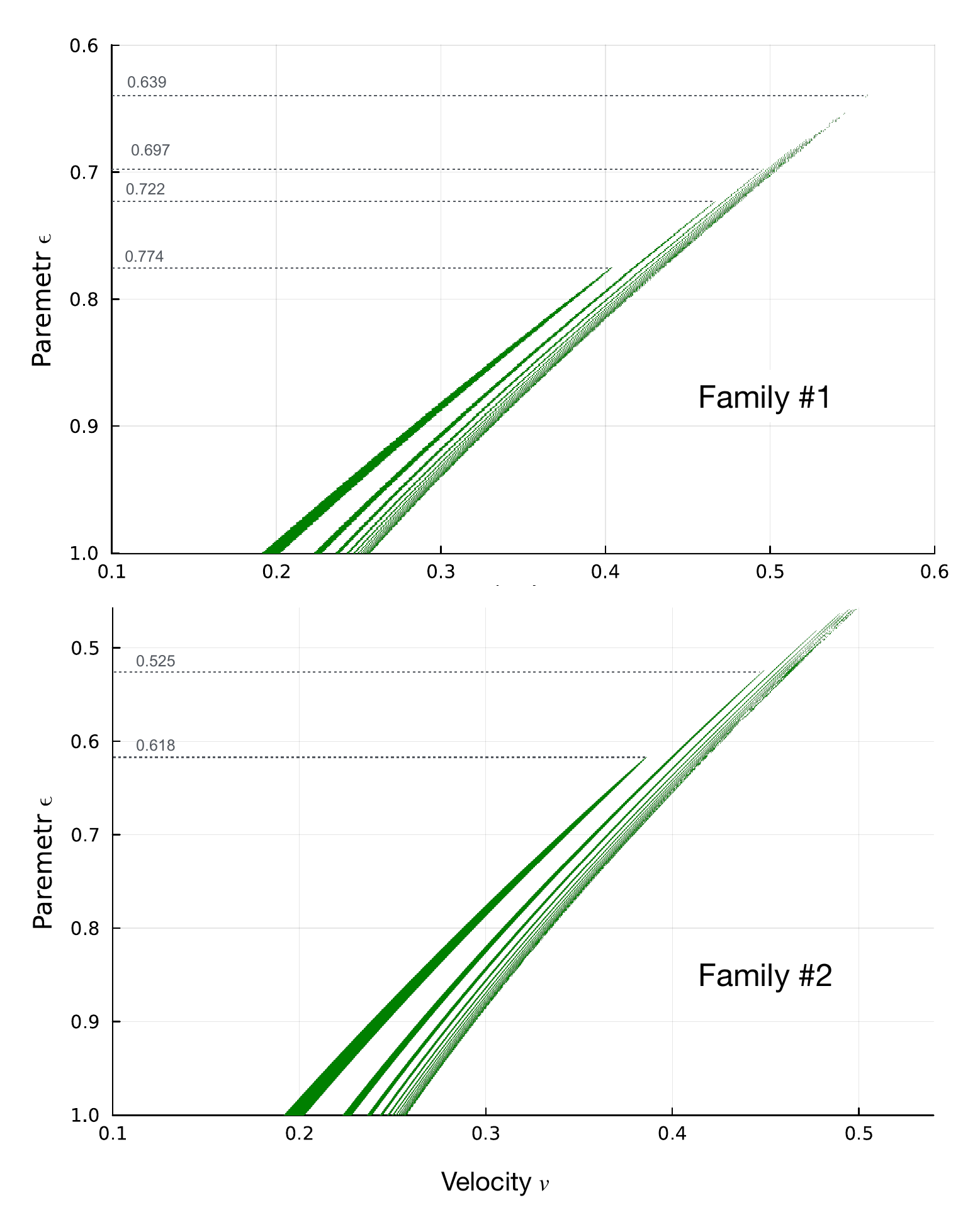}
\end{center}
{
    \caption{\small Closing of two-bounce windows for both families. The numbers represent approximate values of $\varepsilon$ at which that particular window closes, given our numerical accuracy.
 }
    \label{fig:2bouncewindows}
}
\end{figure}

\begin{figure}[htb!]
\begin{center}
\includegraphics[width=0.90\columnwidth]{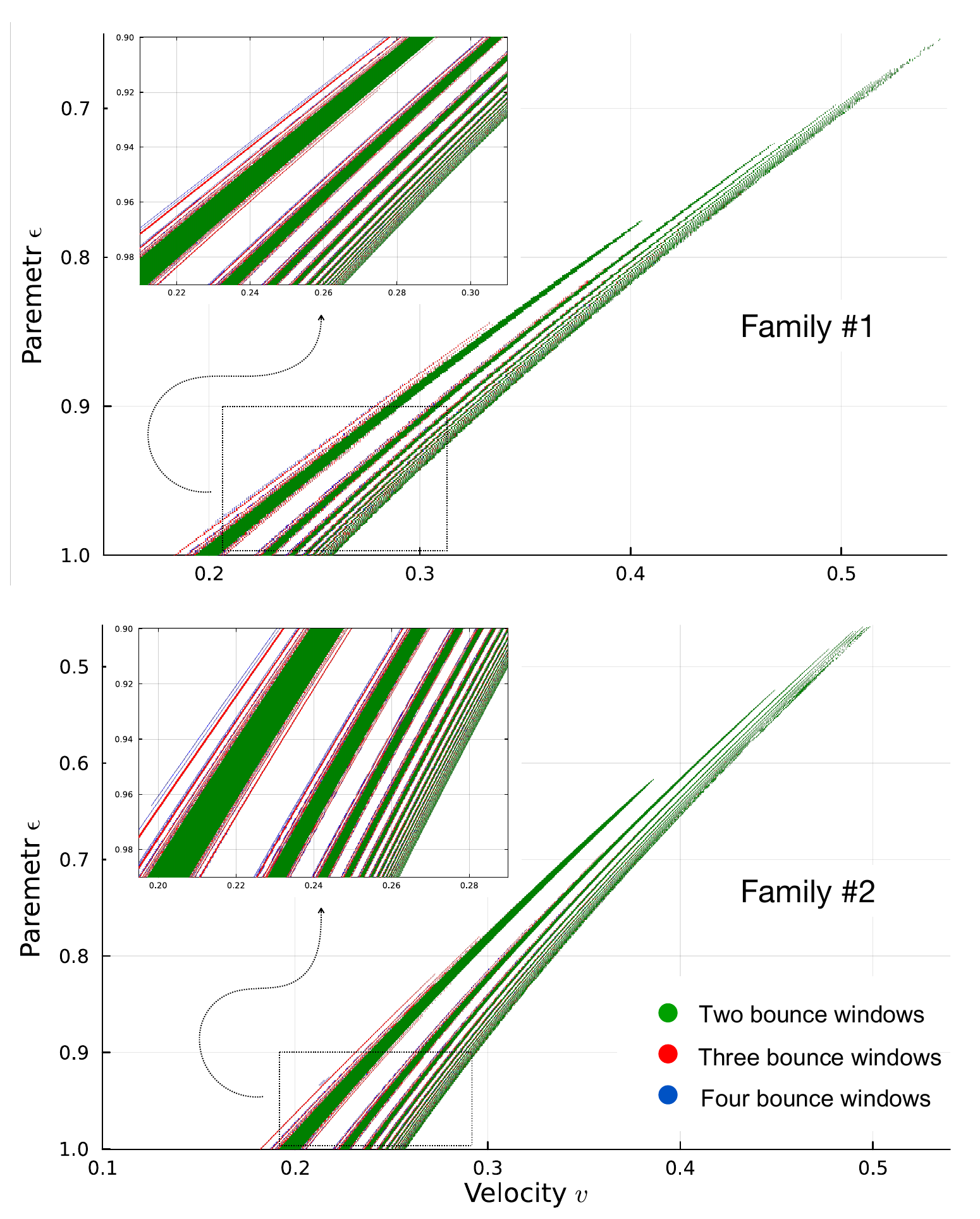}
\end{center}
{
    \caption{\small Detail of closing of windows for two-bounce, three-bounce, and four-bounce for both families. 
 }
    \label{fig:fullbouncewindows}
}
\end{figure}

This observation is reinforced by very similar results seen for the second family (Fig.~\ref{fig:maps} right column). There we see a more gradual closing of the bouncing windows, but they too disappear in spite of the fact that the massive mode is present for all $\varepsilon < 1$. Again, the threshold below which no bouncing windows are observed, i.e. $\varepsilon \approx 0.4$ is quite close to the point of merging of Derrick's mode into the continuum $\varepsilon \approx 0.37$ (see Fig.~\ref{fig:MassDerrick}). 

In Fig.~\ref{fig:2bouncewindows}, we display the range of velocities at which two-bounce windows are found for both families of potentials. There we see that -- perhaps contrary to intuition -- the windows close starting with the largest ones from left to right.  This means, for example, that there exist ranges of $\varepsilon$ for which the two-bounce windows only contain no less than $n$ internal oscillations of the field, with $n = 2, 3, \ldots$.

In Fig.~\ref{fig:fullbouncewindows}, we display a similar picture with added three-bounce and four-bounce windows. The $\varepsilon$ dependence of these windows closely follows the trend set by the two-bounce windows.

\begin{figure}[htb!]
\begin{center}
\includegraphics[width=0.90\columnwidth]{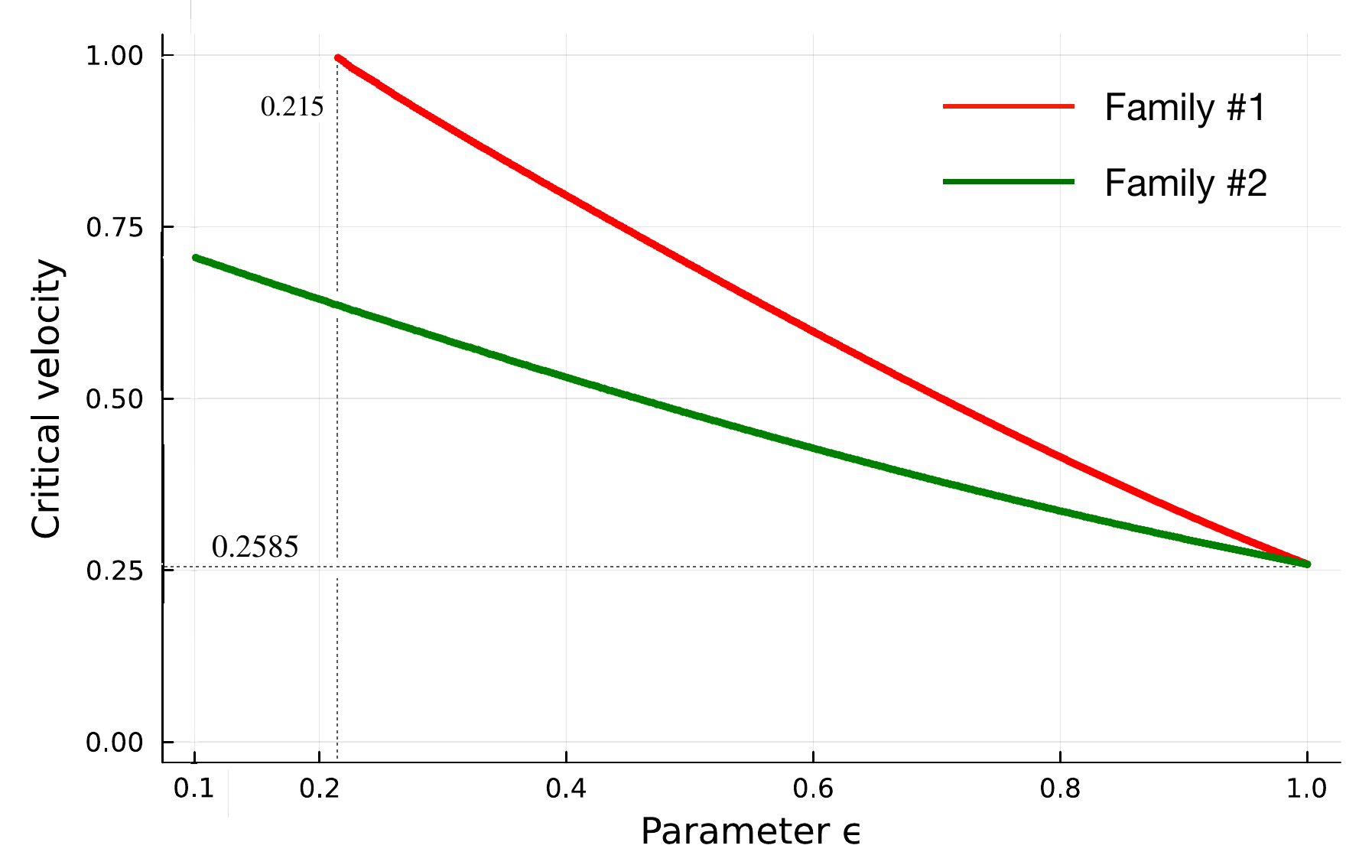}
\end{center}
{
    \caption{\small Dependence of critical velocities on $\varepsilon$ for families $V_{\varepsilon}^{(1,2)}$. For  $V_{\varepsilon}^{(1)}$, the critical velocity crosses the speed of light around $\varepsilon \approx 0.215$. 
 }
    \label{fig:critvels}
}
\end{figure}

Let us also comment on the critical velocity, whose dependence on $\varepsilon$ we display in Fig.~\ref{fig:critvels}. There we see a nearly linear dependence on $\varepsilon$ that is very similar to the dependence of Derrick's frequency on Fig.~\ref{fig:MassDerrick} (bottom).

%%%%%%%%%%%%%%%%%%%%%%%%%%%%%%%%%%%%%%%%%%%%%%%%%%%%%%%%%%
%%%%%%%%%%%%%%%%%%%%%%%%%%%%%%%%%%%%%%%%%%%%%%%%%%%%%%%%%%
%%%%%%%%%%%%%%%%%%%%%%%%%%%%%%%%%%%%%%%%%%%%%%%%%%%%%%%%%%
%%%%%%%%%%%%%%%%%%%%%%%%%%%%%%%%%%%%%%%%%%%%%%%%%%%%%%%%%%
\section{Summary and outlook}
\label{sec:V}
%%%%%%%%%%%%%%%%%%%%%%%%%%%%%%%%%%%%%%%%%%%%%%%%%%%%%%%%%%
%%%%%%%%%%%%%%%%%%%%%%%%%%%%%%%%%%%%%%%%%%%%%%%%%%%%%%%%%%
%%%%%%%%%%%%%%%%%%%%%%%%%%%%%%%%%%%%%%%%%%%%%%%%%%%%%%%%%%
%%%%%%%%%%%%%%%%%%%%%%%%%%%%%%%%%%%%%%%%%%%%%%%%%%%%%%%%%%

In this paper, we have introduced the notions of a core, tails, and skin of a kink -- approximate regions of a generic two-vacuum potential, where the solutions to the leading-order Taylor-expanded equation of motion around the central maximum, minima, and inflection points, respectively, holds. Further, we have introduced a class of potentials, that we call Frankensteinian, as ideal laboratories for exploring the significance of these structural pieces of a kink on the dynamics. In particular, in this paper, we were concerned with the presence or absence of bouncing in $K\bar{K}$ scattering. 

In Sec.~\ref{sec:II}, we have shown three concrete examples of coreless potentials or increasing complexity and we demonstrated a lack of bouncing in all of them. We pointed out that the absence of both core and skin regions results in trivial $K\bar{K}$ scattering outcomes: namely complete annihilation of the pair. Only when the potential had skin regions, do we see oscillons and the appearance of critical velocity.

In Sec.~\ref{sec:III}, we explored the impact of non-analyticity at the maximum for the bouncing phenomenon by considering two parametric families of potentials that are increasingly more singular at $\phi = 0$ as $\varepsilon \to 0$. 

 The first family, $V_{\varepsilon}^{(1)}$, approaches the harmonic double-well TT potential \refer{eq:cusp} as $\varepsilon \to 0$ and, therefore, departs the most from the canonical $\phi^4$ model. Indeed, as $\varepsilon \to 0$, the scattering windows rapidly close, the critical velocity increases almost linearly and all resonant features disappear well before $\varepsilon =0$. 
 
\begin{figure*}[htb!]
\begin{center}
\includegraphics[width=0.84\textwidth]{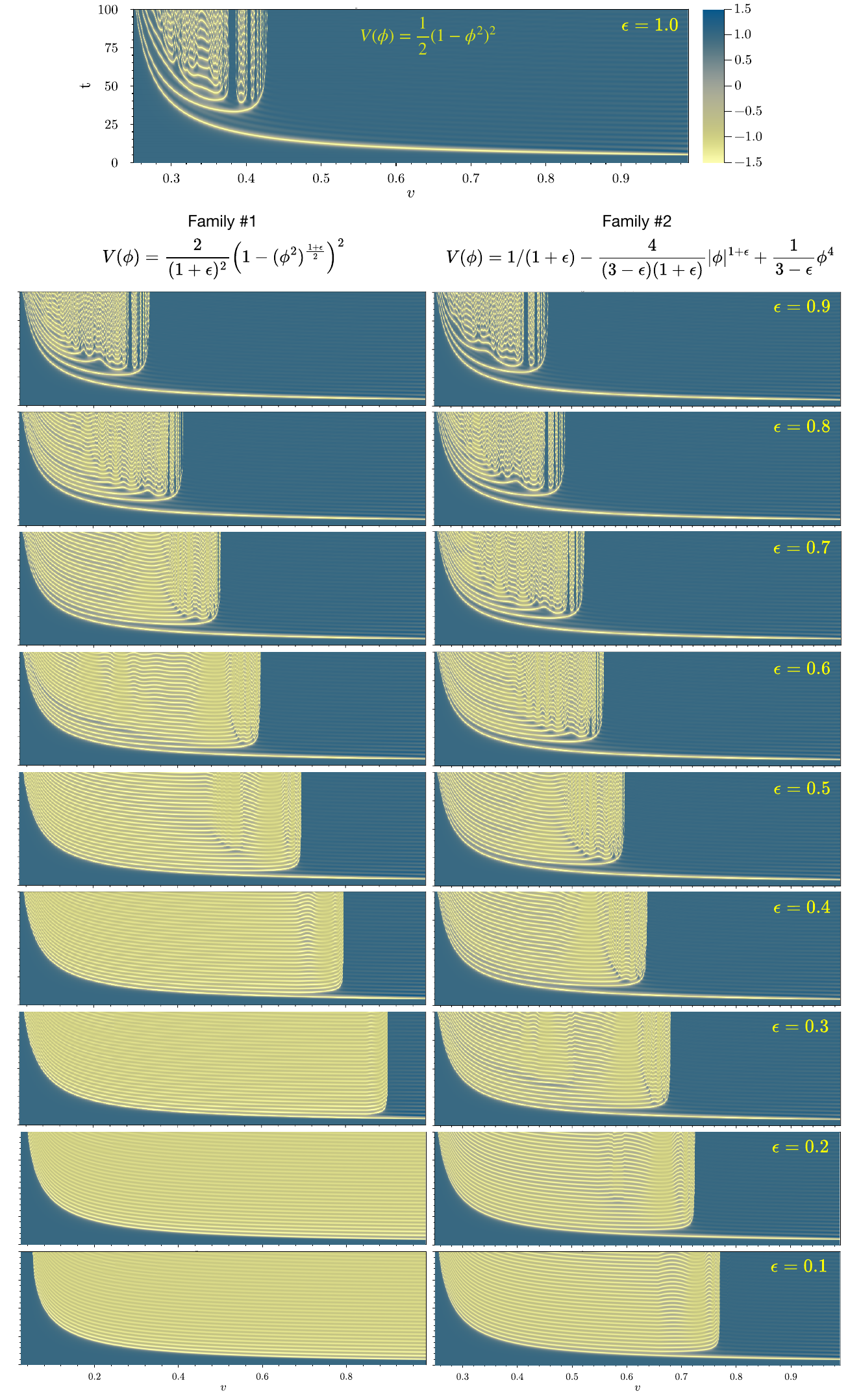}
\end{center}
{
    \caption{\small Gradual disappearance of resonant features in $K\bar{K}$ collisions for $V_{\varepsilon}^{(1)}$ and $V_{\varepsilon}^{(2)}$  as $\varepsilon$ is lowered from $\phi^4$  model ($\varepsilon = 1.0$).
 %The annihilation process completely dominates the $K\bar{K}$ collisions from $\varepsilon = 0.02$ downwards.
 }
    \label{fig:maps}
}
\end{figure*}

 Perhaps surprisingly, the second family which was designed to stay quite close to the $\phi^4$ model for the entire range of $\varepsilon$, showed very similar results. The most visible difference compared with the first family is that the critical velocity remains below one and, as Fig.~\ref{fig:maps} indicates, some oscillons are formed as the outcome of the collision even for very small $\varepsilon$. In that regard, the $\varepsilon \to 0$ limit is qualitatively similar to the $V_{\rm KG+log}$ potential discussed in Sec.~\ref{sec:II}. 
 
Perhaps most surprisingly, in both families, we have seen that the instance of $\varepsilon$ when Derrick's frequency merges with the continuum (see Fig.~\ref{fig:MassDerrick}) is a far superior predictor of when the bouncing windows disappear compared with the massive mode. This is certainly unexpected, but let us speculate on why this could be the case. 

As $\varepsilon \to 0$, the range of initial velocities containing bouncing windows gets quickly shifted towards higher, relativistic values. It is known that Derrick's mode plays a key role in relativistic dynamics, hence it is also important here. To test this hypothesis, it would be of interest to find a family of models for which critical velocity remains small for the entire range of the deformation parameter.
 
Our results suggest that the core region of the potential plays a central (pun intended) role in the bouncing phenomenon in $K\bar{K}$ scattering. Of course, it is hard to quantify this statement, and, at this point, we do not have sufficient knowledge to turn the observations of this paper into a predictive heuristic. Nevertheless, we feel that the presented results give credence to the notion that core, tails, and skins, as defined above, are of some importance to the dynamical phenomena and that they merit further investigation.

In particular, we plan to present a detailed investigation of lifetimes and other characteristics of oscillons in a separate publication. 
We also plan to investigate further examples of Frankensteinian potentials and, in particular, determine the role of the skin on the dynamics of kinks.

\acknowledgments

We would like to thank T.~Roma\'nczukiewicz, K. S\l awi\' nska, and A.~Wereszczy\'nski for many fruitful discussions and hospitality at Jagiellonian University in Krakow. This work benefited greatly from the feedback provided by the participants of the SIG XII workshop held at Jagiellonian University in June 2024. 
We also want to say a huge thanks to Martin Urbanec who helped us immensely with running the numerical code. 
We also acknowledge the institutional support of the Research Centre for Theoretical Physics and Astrophysics, Institute of Physics, Silesian University in Opava.
This work has been supported by the grant no. SGS/24/2024 Astrophysical processes in strong gravitational and electromagnetic fields of compact object.

\appendix

\section*{Appendix: Comments on the numerical method}
\label{app:A}

All numerical calculations were performed in the programming language Julia 1.10.3 using the library DifferentialEquations.jl. We evolved the second-order equations of motion via the Bogacki-Shampine 5/4 Runge-Kutta method (BS5()) in the time domain, while the spatial domain was discretized (typically into $\sim$6000 segments), approximating the second derivative using the central finite difference of the third-order.

We also exploited the reflection symmetry through the center of collision at $x=0$, so that the calculations were performed only in the half interval $x \in \lbrack 0,x_{end} \rbrack$.
For the last 10\% of the length of the interval, we introduced an absorbing layer in which an additional damping term is included in the equation of motion,i.e.
\begin{equation*}
\partial_t ^2 \phi=  \partial_x ^2 \phi - V^\prime(\phi) - \alpha_D  \partial_t\phi\,,
\end{equation*}
where the damping value $\alpha_D$ was chosen in the range of values  $\alpha_D \in \lbrack 0.2,0.4 \rbrack$ depending on the complexity of  the numerical calculation. This approach allows us to use a smaller spatial interval $x \in \lbrack 0,25 \rbrack$ without fear of reflections from the end of the interval.

%%%%%%%%%%%%%%%%

\end{document}